\documentclass[aip,jmp,notitlepage,showpacs,showkeys]{revtex4-1}
\usepackage{amsmath,amssymb,amsthm}
\usepackage[pdftoolbar=true,pdfmenubar=false,bookmarks=true,colorlinks=false]{hyperref}
\newtheorem{thm}{Theorem}
\theoremstyle{definition}
\newtheorem{definition}{Definition}
\newtheorem{example}{Example}
\begin{document}

\title{Cabibbo-Kobayashi-Maskawa matrix:  parameterizations and rephasing invariants}

\author{H. P\'erez R.}

\thanks{Work was partly supported by Polish Ministry of Science and Higher Education Grant N~N202~230337}

\author{P. Kielanowski}

\email{kiel@fis.cinvestav.mx}

\affiliation{Departamento de F\'{\i}sica, Centro de Investigaci\'on y de Estudios Avanzados,
07000 M\'exico D.F., Mexico}

\author{S.R. Ju\'arez W.}

\thanks{Proyecto SIP: 20120639, Secretar\'{\i}a de Investigaci\'on y
  Posgrado, Beca EDI y Comisi\'on de Operaci\'on y Fomento de
  Actividades Acad\'emicas (COFAA) del IPN}
\affiliation{Departamento de F\'{\i}sica, Escuela Superior de   F\'{\i}sica y Matem\'{a}ticas, Instituto Polit\'{e}cnico Nacional,  U.P. ``Adolfo L\'{o}pez Mateos''. C.P.~07738, M\'{e}xico D.F.}

\email{rebeca@esfm.ipn.mx}

\begin{abstract}
The paper is devoted to a discussion of general properties of the Cabibbo-Kobayashi-Maskawa (CKM) matrix. First we propose a general me\-thod of a recursive construction of the CKM matrix for any number of generations. This allows to set up a parameterization with desired properties. As an application we generalize the Wolfenstein parameterization to the case of 4~generations and obtain restrictions on the CKM suppression of the fourth generation. Motivated by the rephasing invariance of the CKM observables we next consider the general phase invariant monomials built out of the CKM matrix elements and their conjugates. We show, that there exist 30 fundamental phase invariant monomials and 18~of them are a product of 4~CKM matrix elements and 12~are a product of 6~CKM matrix elements. In the Main Theorem we show that all rephasing invariant monomials can be expressed as a product of at most 5~factors: 4~of them are fundamental phase invariant monomials and the fifth factor consists of  powers of squares of absolute values of the CKM matrix elements.
\end{abstract}


\maketitle
\section{Introduction}\label{sec:1}
The Standard Model has 18~adjustable parameters and 13~of these parameters have their origin in the Yukawa couplings of the Higgs boson with fermions. These parameters are: 3~lepton masses, 6~quark masses and 4~parameters of the CKM matrix~\cite{M.-Kobayashi:1973qy}. The quark masses are the running masses and they are obtained from the eigenvalues of the Yukawa couplings in the process of diagonalization by the bi-unitary transformation. The CKM matrix is obtained from the \textit{left} diagonalizing matrices of the up- and down quark mass matrices. The CKM matrix is a $3\times3$ unitary matrix, which has additional properties stemming from the rephasing freedom of the quark fields. This is the reason why the CKM matrix has only 4~real parameters and one of them is a phase. The non vanishing phase is the source of the CP~violation in the Standard Model. 

The CKM matrix has 9 complex matrix elements, which are parameterized by 4~real parameters. The choice of the parameters is not unique. There exist  various equivalent parameterizations, which were chosen to fulfill various needs. Let us start with the standard parameterization of the PDG~\cite{0954-3899-37-7A-075021}. This parameterization is exact and uses 3~angles and 1~phase and can be represented as the product of 3~real rotation matrices and the diagonal matrices with phase terms.

Another widely used parameterization is the one proposed by Wolfenstein~\cite{PhysRevLett.51.1945}. Initially it was considered to be an approximate representation of the CKM matrix, because it was chosen in such a way as to reproduce the suppression for the weak transitions of quarks between the generations. Later, it was improved~\cite{PhysRevD.50.3433,PhysRevD.51.3958,PhysRevD.33.860} and also it was made exact~\cite{PTP.92.287}.

The particular choice of parameters in principle is not important, because all exact parameterizations are mathematically equivalent. But an appropriate choice of parameterization for a given physical situation can lead to natural relationship between the parameters and physical observables. Fo example, in the parameterization of Wolfenstein the Jarlskog's invariant~\cite{PhysRevLett.55.1039} $J=A\eta\lambda^{6}+\mathcal{O}(\lambda^{8})$ has simple form, while in the standard parameterization of the CKM matrix~\cite{0954-3899-37-7A-075021} it is more complicated.

In Section~\ref{sec:2} we discuss the recursive construction of the CKM matrix for a general case of $n$~generations, assuming that the CKM matrix for $(n-1)$~quark generations is known. There exist various approaches to the problem of parameterization of the CKM matrix, which fulfill various requirements. The parameterization of Harari and Leuler~\cite{Harari1986123} stresses the simplicity and a possibility of generalization to an arbitrary number of quark generations. Jarlskog proposed the recursive parameterization also for an arbitrary number of quark generations that represents the CKM matrix as a product of special unitary matrices~\cite{Jarlskog:2005zb,Jarlskog:2005zw}. Bjorken and Dunietz~\cite{Bjorken:1987tr} construct the CKM matrix out of rephasing invariants: the absolute values of the CKM matrix elements $\lvert V_{i\alpha}\rvert$, $i<\alpha$ and the phase of ``plaquette'' $\operatorname{arg}(V_{i\alpha}V_{j\beta}V_{i\beta}^{*}V_{j\alpha}^{*})$. There are many more parameterizations~\cite{Chaturvedi:2010qf,springerlink:10.1007/BF02099133,PhysRevLett.53.1802,springerlink:10.1007/BF02099985,PhysRevD.35.1732,PhysRevD.57.594,Gupta:2001,PhysRevLett.63.2189,Maiani1976183,PhysRevD.59.093009,springerlink:10.1007/BF02734434,PhysRevLett.61.35} which help to illustrate various features of the CKM matrix, but their discussion is beyond the scope of our paper.

Our algorithm for the construction of the parameterization of the CKM matrix has the flexibility that allows to adjust the parameterization to the required physical situation. It can be used to generalize the standard and Wolfenstein parameterizations to the case of 4~quark generations. From the Wolfenstein-type parameterization for 4~generations we then obtain the restrictions on the suppression factors for the 4-th generation, which are compatible with the present data for the CKM matrix.

The next topic, discussed in Section~\ref{sec:3}, is the quark fields rephasing properties of the CKM matrix. The Yukawa couplings (mass matrices) of the standard model are not invariant under the rephasing of the quark fields. From this it follows that the CKM matrix is not invariant either. Since observable effects cannot depend on the choice of the phase of the quark fields it means that all observables including those containing the CKM matrix must not depend on arbitrary phases of the quark fields. This is the reason of the reduction of the number of the physically significant parameters of the unitary CKM matrix from 9~to~4.

The general observables  of CKM matrix are usually monomials built out of the CKM matrix elements $V_{ij}$ and its conjugates $V_{ij}^{*}$. The most important observables obtained from the rephasing invariant monomials of the CKM matrix\footnote{The squares of the absolute values of the CKM matrix elements $\lvert V_{ij}\rvert^{2}$ are clearly rephasing invariant observables, but we are focused on observables sensitive to the phases of the CKM matrix elements.} are the Jarlskog invariant~\cite{PhysRevLett.55.1039}~~\footnote{Also the following formula holds
\[
\operatorname{Im}(V_{\alpha i}V_{\beta j}V_{\alpha j}^{*}V_{\beta i}^{*}) =J\sum_{\gamma,k}\varepsilon_{\alpha\beta\gamma}\varepsilon_{ijk} =J(3(\delta_{\alpha i}\delta_{\beta j}-\delta_{\alpha j}\delta_{\beta i}) +\delta_{\alpha i}+\delta_{\beta j}-\delta_{\alpha j}-\delta_{\beta i}).
\]} $J$
\begin{equation}
\label{eq:I.1}
J=\operatorname{Im}(V_{11}V_{22}V_{12}^{*}V_{21}^{*}),
\end{equation}
and the unitarity triangle angles:
\begin{equation}
\label{eq:I.2}
\begin{split}
&\alpha=\operatorname{arg}\left (-
\frac{V_{31}V_{33}^{*}}{V_{11}V_{13}^{*}}
\right )=
\operatorname{arg}\left (-
V_{13}V_{31}V_{11}^{*}V_{33}^{*}
\right ),\\
&\beta=\operatorname{arg}\left (-
\frac{V_{21}V_{23}^{*}}{V_{31}V_{33}^{*}}
\right )=
\operatorname{arg}\left (-
V_{21}V_{33}V_{23}^{*}V_{31}^{*}
\right ),\\
&\gamma=\operatorname{arg}\left (-
\frac{V_{11}V_{13}^{*}}{V_{21}V_{23}^{*}}
\right )=
\operatorname{arg}\left (-
V_{11}V_{23}V_{13}^{*}V_{21}^{*}.
\right )
\end{split}
\end{equation}

In Section~\ref{sec:3} we derive the conditions for the rephasing invariance of an arbitrary monomial built from the CKM matrix elements and then demonstrate in the Main Theorem that an arbitrary rephasing monomial can be expressed as the product of five factors: powers of absolute values of the CKM matrix elements multiplied by powers of 4~fundamental rephasing monomials, which are built out of 4~and 6 CKM matrix elements.

The last section of the paper contains a summary and conclusions.

\section{Parameterizations of the CKM matrix}\label{sec:2}

\subsection{Introductory remarks}\label{sec:2.1}

The CKM matrix is unitary, but the rephasing freedom for the quark
fields reduces the number of the physically significant
parameters. The $(n\times n)$ unitary matrix is described by $n^{2}$
parameters. The up- and down-quarks rephasing freedom reduces the
number of the parameters by $(2n-1)$, so the CKM matrix for $n$ quark
generations is described by $(n-1)^{2}$ parameters. These parameters
are divided into two classes: angle-like and phases. Angle-like parameters are derived from the $(n\times n)$ \emph{real} unitary matrix~\footnote{For example the parameter $\lambda$ for the $2\times 2$ rotation matrix
\[
\left( 
\begin{array}{cc}
\sqrt{1-\lambda^{2}}, &\lambda\\
-\lambda,&\sqrt{1-\lambda^{2}}
\end{array}
\right)
\]
will be called the angle-like variable.} (rotation or orthogonal
matrix) and there are $\frac{n(n-1)}{2}$ such parameters. The remaining $\frac{(n-1)(n-2)}{2}$ parameters are phases. One can observe that if
the number of quark generations is incremented from $(n-1)$ to $n$
then the number of angle-like parameters increases by $(n-1)$ and the number of
phases by $(n-2)$.

We will present here the recursive construction of the $(n\times n)$
CKM matrix $V^{(n)}$, assuming that the $(n-1)\times(n-1)$ CKM matrix
$V^{(n-1)}$ is known. Let us introduce the notation, where the
parameters of the CKM matrix (angle-like and phases) are labeled
according to the generation number to which they belong:
\begin{equation}
\label{eq:II.1}
\begin{array}{l}
\theta_{1,k},\theta_{2,k},\ldots,\theta_{k-1,k}\text{ - angle-like parameters for the
  $k$-th generation}\\
\delta_{1,k},\delta_{2,k},\ldots,\delta_{k-2,k}\text{ - phases for the
  $k$-th generation}
\end{array}
\end{equation}
In such a way the following hierarchy of the parameters has been formed:
\begin{equation}
\label{eq:II.2}
\begin{array}{c|l}
\text{generation}&\text{parameters}\\
\hline
2&\theta_{1,2}\\
3&\theta_{1,3},\theta_{2,3},\delta_{1,3}\\
4&\theta_{1,4},\theta_{2,4},\theta_{3,4},\delta_{1,4},\delta_{2,4}\\
\cdots&\cdots
\end{array}.
\end{equation}
The continuation of the Table in Eq.~\eqref{eq:II.2} is
obvious. The $(n\times n)$ CKM matrix contains the parameters from the
Table in Eq.~\eqref{eq:II.2} that are in the $n$-th row and above.

\subsection{Recursive construction of the CKM matrix}\label{sec:2.2}

In this section we outline how to construct and parameterize the $(n\times n)$ CKM
matrix $V^{(n)}$, if we know the CKM matrix $V^{(n-1)}$. The presented algorithm does not impose any conditions on the parameterization of the matrix $V^{(n-1)}$, so this method allows to introduce such properties of the CKM matrix that are required by the physical conditions for each generation separately. 

Let us first introduce the necessary notation. We write the CKM
matrix $V^{(n)}$ in terms of column vectors
\begin{equation}
\label{eq:II.3}
V^{(n)}=\left(\mathbf{v}_{1}^{(n)},\mathbf{v}_{2}^{(n)},\ldots,
  \mathbf{v}_{n}^{(n)}\right)
\end{equation}
i.e., the vectors $\mathbf{v}_{1}^{(n)},\mathbf{v}_{2}^{(n)},\ldots,
\mathbf{v}_{n}^{(n)}$ are constructed from the elements of the matrix
$V^{(n)}$
\begin{equation}
\label{eq:II.3a}
\mathbf{v}_{k}^{(n)}=\left (
\begin{array}{c}
V_{1k}^{(n)}\\
V_{2k}^{(n)}\\
\vdots\\
V_{nk}^{(n)}
\end{array}
\right ).
\end{equation}

We assume that the explicit form of the matrix $V^{(n-1)}$ is known, so
according to Eqs.~\eqref{eq:II.1} and~\eqref{eq:II.2}  it is a function of the following
parameters:
\[
\begin{array}{l}
\theta_{1,2},\theta_{1,3},\theta_{2,3},\ldots,\theta_{n-2,n-1},\\
\delta_{1,3},\delta_{1,4},\delta_{2,4},\ldots,\delta_{n-3,n-1}
\end{array}.
\]
The CKM matrix $V^{(n)}$ is built from $V^{(n-1)}$ in two steps:
\begin{enumerate}
\item[a.] We construct $n$ \emph{real} column vectors with $n$ rows
\begin{equation}
\label{eq:II.4}
\mathbf{e}_{1},\ldots,\mathbf{e}_{n}
\end{equation}
that depend on $(n-1)$ angle-like independent parameters $\theta_{1,n},\ldots,\theta_{n-1,n}$,
are normalized to~1 and orthogonal:
\begin{equation}
\label{eq:II.4a}
\mathbf{e}_{i}\cdot\mathbf{e}_{j}=\delta_{ij}.
\end{equation}
\item[b.] The columns of the CKM matrix $V^{(n)}$ are obtained from
  the vectors $\mathbf{e}_{k}$ and the elements of the matrix
  $V_{ij}^{(n-1)}$ in the following way
\begin{equation}
\label{eq:II.5}
\begin{array}{l}
  \mathbf{v}_{k}^{(n)}=V_{1k}^{(n-1)}\mathbf{e}_{1} +\sum_{l=2}^{n-1}
  V_{lk}^{(n-1)}\text{e}^{-i\delta_{l-1,n}}\mathbf{e}_{l},\quad k=1,\ldots,n-1,\\
  \mathbf{v}_{n}^{(n)}=\mathbf{e}_{n},
\end{array}\quad n\geq 2
\end{equation}
\end{enumerate}
and this completes the construction of the CKM matrix $V^{(n)}$ (see
Eq.~\eqref{eq:II.3}).
The matrix $V^{(n)}$ constructed in such a way has following
properties
\begin{enumerate}
\item[a.] It is unitary. This follows from the unitarity of the matrix
  $V^{(n-1)}$ and the orthogonality~\eqref{eq:II.4a} of the vectors
  $\mathbf{e}_{l}$.
\item[b.] It depends on parameters of the matrix $V^{(n-1)}$ and on $(n-1)$ parameters of the vectors $\mathbf{e}_{1},\ldots,\mathbf{e}_{n}$ and on $(n-2)$ phase factors $\text{e}^{-i\delta_{l,n}}$ from~\eqref{eq:II.5}. Altogether it depends on $\frac{n(n-1)}{2}$ angle-like variables and $\frac{(n-1)(n-2)}{2}$ phases.
\end{enumerate}

The resulting parameterization of the matrix $V^{(n)}$ depends on the
pa\-ra\-me\-tri\-zation of $V^{(n-1)}$ and that of the vectors
$\mathbf{e}_{1},\ldots,\mathbf{e}_{n}$ and on the phase factors $\text{e}^{-i\delta_{l,n}}$. 
If we additionally assume that the vectors $\mathbf{e}_{i}$ fulfill the conditions
\begin{equation}
\label{eq:II.4aa}
\left. (\mathbf{e}_{i})_{j}\right\rvert_{\substack{\theta_{k,n}=0\\ \delta_{l,n}=0}}
=
\begin{cases}
1&\text{if }i=j\\
0&\text{otherwise}
\end{cases}\quad
\begin{array}{l}
k=1,\ldots,n-1\\
l=1,\ldots,n-2.
\end{array}
\end{equation}
then one obtains
\begin{equation}
\label{eq:II.6A}
\left. V^{(n)}\right\rvert_{\substack{\theta_{k,n}=0\\ \delta_{l,n}=0}}
=\left( \begin{array}{cc}
    V^{(n-1)}&0\\
    0&1
\end{array}\right)\quad
\begin{array}{l}
k=1,\ldots,n-1\\
l=1,\ldots,n-2.
\end{array}
\end{equation}

\begin{example} \textit{Standard parameterization for 3 quark generations}\\
We will show here how one can obtain the CKM matrix $V^{(3)}$ in the
standard parameterization using the procedure outlined above.

The matrix $V^{(2)}$ depends on one angle $\theta_{1,2}$ (it is the
$2\times 2$ rotation matrix)
\begin{equation}
\label{eq:II.7}
V^{(2)}=
\left(
\begin{array}{cc}
\cos\theta_{1,2}&\sin\theta_{1,2}\\
-\sin\theta_{1,2}&\cos\theta_{1,2}
\end{array}
 \right)\equiv
\left(
\begin{array}{cc}
c_{1,2}&s_{1,2}\\
-s_{1,2}&c_{1,2}
\end{array}
 \right),
\end{equation}
where we use the notation $s_{i,j}=\sin\theta_{i,j}$ and
$c_{i,j}=\cos\theta_{i,j}$.  We now choose the vectors
$\mathbf{e}_{1}$, $\mathbf{e}_{2}$ and $\mathbf{e}_{3}$ in the
following way
\begin{equation}
\label{eq:II.8}
\mathbf{e}_{1}=\left( 
\begin{array}{c}
c_{1,3}\\
-s_{1,3}s_{2,3}\\
-s_{1,3}c_{2,3}
\end{array}\right), \;
\mathbf{e}_{2}=\left( 
\begin{array}{c}
  0\\
  c_{2,3}\\
  -s_{2,3}
\end{array}
\right),\;
\mathbf{e}_{3}=\left( 
\begin{array}{c}
s_{1,3}\\
c_{1,3}s_{2,3}\\
c_{1,3}c_{2,3}
\end{array}\right) .
\end{equation}
The vectors in Eq.~\eqref{eq:II.8} fulfill the
condition~\eqref{eq:II.4a}.

Now, according to Eq.~\eqref{eq:II.5} we construct the columns
of the matrix $V^{(3)}$
\begin{subequations}\label{eq:II.9}
\begin{equation}
\label{eq:II.9a}
\mathbf{v}_{1}^{(3)}=V_{11}\mathbf{e}_{1}+V_{21}\text{e}^{-i\delta_{1,3}}\mathbf{e}_{2}=
\left(
\begin{array}{c}
c_{1,2}c_{1,3}\\
-c_{1,2}s_{1,3}s_{2,3}-\text{e}^{-i\delta_{1,3}}s_{1,2}c_{2,3}\\
-c_{1,2}s_{1,3}c_{2,3}+\text{e}^{-i\delta_{1,3}}s_{1,2}s_{2,3}
\end{array}
 \right),
\end{equation}
\begin{equation}
\label{eq:II.9b}
\mathbf{v}_{2}^{(3)}=V_{12}\mathbf{e}_{1}+V_{22}\text{e}^{-i\delta_{1,3}}\mathbf{e}_{2}=
\left(
\begin{array}{c}
s_{1,2}c_{1,3}\\
-s_{1,2}s_{1,3}s_{2,3}+\text{e}^{-i\delta_{1,3}}c_{1,2}c_{2,3}\\
-s_{1,2}s_{1,3}c_{2,3}-\text{e}^{-i\delta_{1,3}}c_{1,2}s_{2,3}
\end{array}
 \right),
\end{equation}
\begin{equation}
\label{eq:II.9c}
\mathbf{v}_{3}^{(3)}=\mathbf{e}_{3}=
\left(
\begin{array}{c}
s_{1,3}\\
c_{1,3}s_{2,3}\\
c_{1,3}c_{2,3}
\end{array}
 \right),
\end{equation}
\end{subequations}
so the matrix $V^{(3)}$ is equal
\begin{equation}\label{eq:II.10}
V^{(3)}=(\mathbf{v}_{1},\mathbf{v}_{3},\mathbf{v}_{3})\\
=\left( 
\begin{array}{ccc}
c_{1,2}c_{1,3}&s_{1,2}c_{1,3}&s_{1,3}\\
-c_{1,2}s_{1,3}s_{2,3}-\text{e}^{-i\delta_{1,3}}s_{1,2}c_{2,3}&
-s_{1,2}s_{1,3}s_{2,3}+\text{e}^{-i\delta_{1,3}}c_{1,2}c_{2,3}&
c_{1,3}s_{2,3}\\
-c_{1,2}s_{1,3}c_{2,3}+\text{e}^{-i\delta_{1,3}}s_{1,2}s_{2,3}&
-s_{1,2}s_{1,3}c_{2,3}-\text{e}^{-i\delta_{1,3}}c_{1,2}s_{2,3}&
c_{1,3}c_{2,3}
\end{array}
\right).
\end{equation}
The form of the matrix~\eqref{eq:II.10} is not exactly the same as
that of the standard parameterization~\cite{0954-3899-37-7A-075021}, but both forms are
equivalent, because after rephasing of the matrix $V^{(3)}$ in
Eq.~\eqref{eq:II.10} by multiplying the first and second column by
$\text{e}^{i\delta_{1,3}}$ and the first row by
$\text{e}^{-i\delta_{1,3}}$ one obtains exactly the standard
parameterization.
\end{example}

\begin{example} \textit{Standard parameterization for 4 quark generations}\\
In this example we will construct an analogue of the standard
parameterization for 4~generations using the standard form of the CKM
matrix~$V^{(3)}$ from Ref.~\cite{0954-3899-37-7A-075021}. First we construct the basis vectors
$\mathbf{e}_{i}$ for the 4-dimensional case
\begin{equation}
\label{eq:II.11}
\begin{array}{llll}
\mathbf{e}_{1}=\left( 
\begin{array}{c}
c_{1,4}\\
-s_{1,4}s_{2,4}\\
-s_{1,4}c_{2,4}s_{3,4}\\
-s_{1,4}c_{2,4}c_{3,4}
\end{array}\right),

&\mathbf{e}_{2}=\left( 
\begin{array}{c}
  0\\
  c_{2,4}\\
  -s_{2,4}s_{3.4}\\
  -s_{2,4}c_{3.4}
\end{array}
\right),

&\mathbf{e}_{3}=\left( 
\begin{array}{c}
0\\
0\\
c_{3,4}\\
-s_{3,4}
\end{array}\right),

&\mathbf{e}_{4}=\left( 
\begin{array}{c}
s_{1,4}\\
c_{1,4}s_{2,4}\\
c_{1,4}c_{2,4}s_{3,4}\\
c_{1,4}c_{2,4}c_{3,4}
\end{array}\right) .

\end{array}
\end{equation}
Then, using Eqs.~\eqref{eq:II.5} we construct the vectors
$\mathbf{v}_{i}^{(4)}$ and the matrix $V^{(4)}$, which we call the
standard parameterization of the CKM matrix for 4~generations. The
explicit form of the matrix elements of $V^{(4)}_{ij}$ is rather lengthy and reads
\begin{equation}\label{eq:II.CKM4}
\begin{split}
V^{(4)}_{11}=&c_{1,2} c_{1,3} c_{1,4},\\
V^{(4)}_{21}=& -e^{-i \delta _{1,4}} c_{2,3} c_{2,4} s_{1,2}-e^{i (\delta _{1,3}- \delta _{1,4
})} c_{1,2} c_{2,4} s_{1,3} s_{2,3}-c_{1,2} c_{1,3} s_{1,4}
   s_{2,4},\\
V^{(4)}_{31}=&-e^{i (\delta _{1,3}- \delta _{2,4})} c_{1,2} c_{2,3} c_{3,4} s_{1,3}+e^{i (\delta _{1,3}- \delta _{1,4})} c_{1,2} s_{2,3} s_{2,4} s_{3,4}
   s_{1,3}\\
  & +e^{-i \delta _{2,4}} c_{3,4} s_{1,2} s_{2,3}-c_{1,2} c_{1,3} c_{2,4
} s_{1,4} s_{3,4}+e^{-i \delta _{1,4}} c_{2,3} s_{1,2} s_{2,4}
   s_{3,4},\\
V^{(4)}_{41}=&-c_{1,2} c_{1,3} c_{2,4} c_{3,4} s_{1,4}+e^{-i \delta _{1,4}} c_{2,3} c_{3,4} s
_{1,2} s_{2,4}-e^{-i \delta _{2,4}} s_{1,2} s_{2,3}
   s_{3,4}\\
   &+e^{i (\delta _{1,3}- \delta _{1,4})} c_{1,2}
   c_{3,4} s_{1,3} s_{2,3} s_{2,4}+e^{i (\delta _{1,3}- \delta _{2,4})} c_{1,2} c
_{2,3} s_{1,3} s_{3,4},\\
V^{(4)}_{12}=&c_{1,3} c_{1,4} s_{1,2},\\
V^{(4)}_{22}=&e^{-i \delta _{1,4}} c_{1,2} c_{2,3} c_{2,4}-e^{i (\delta _{1,3}-\delta _{1,4})
} s_{1,2} s_{1,3} s_{2,3} c_{2,4}-c_{1,3} s_{1,2} s_{1,4}
   s_{2,4},\\
V^{(4)}_{32}=& -e^{i (\delta _{1,3}- \delta _{2,4})} c_{2,3} c_{3,4} s_{1,2} s_{1,3}+e^{i (\delta _{1,3}-\delta _{1,4})} s_{1,2} s_{2,3} s_{2,4} s_{3,4}
   s_{1,3}\\
   &-e^{-i \delta _{2,4}} c_{1,2} c_{3,4} s_{2,3}-c_{1,3} c_{2,4} s_{1,2
} s_{1,4} s_{3,4}-e^{-i \delta _{1,4}} c_{1,2} c_{2,3} s_{2,4}
   s_{3,4},\\
V^{(4)}_{42}=&-c_{1,3} c_{2,4} c_{3,4} s_{1,2} s_{1,4}-e^{-i \delta _{1,4}} c_{1,2} c_{2,3} c
_{3,4} s_{2,4}+e^{-i \delta _{2,4}} c_{1,2} s_{2,3}s_{3,4}\\
&+e^{i (\delta _{1,3}- \delta _{1,4})} c_{3,4}
   s_{1,2} s_{1,3} s_{2,3} s_{2,4}+e^{i (\delta _{1,3}-\delta _{2,4})} c_{2,3} s
_{1,2} s_{1,3} s_{3,4},\\
V^{(4)}_{13}=&e^{-i \delta _{1,3}} c_{1,4} s_{1,3},\\
V^{(4)}_{23}=&e^{-i \delta _{1,4}} c_{1,3} c_{2,4} s_{2,3}-e^{-i \delta _{1,3}} s_{1,3} s_{1,
4} s_{2,4},\\
V^{(4)}_{33}=&e^{-i \delta _{2,4}} c_{1,3} c_{2,3} c_{3,4}-e^{-i \delta _{1,3}} c_{2,4} s_{1,
3} s_{1,4} s_{3,4}-e^{-i \delta _{1,4}} c_{1,3} s_{2,3}
   s_{2,4} s_{3,4},\\
V^{(4)}_{43}=&-e^{-i \delta _{1,3}} c_{2,4} c_{3,4} s_{1,3} s_{1,4}-e^{-i \delta _{1,4}} c_{1
,3} c_{3,4} s_{2,3} s_{2,4}-e^{-i \delta _{2,4}} c_{1,3}
   c_{2,3} s_{3,4},\\
\\
V^{(4)}_{14}=&s_{14},\quad
V^{(4)}_{24}=c_{1,4}s_{2,4},\quad
V^{(4)}_{34}=c_{1,4}c_{2,4}s_{3,4},\quad
V^{(4)}_{44}=c_{1,4}c_{2,4}c_{3,4}.
\end{split}
\end{equation}
\end{example}

\subsection{Wolfenstein parameterization for 4~quark generations}\label{sec2.3}

The other widely used parameterization of the CKM matrix is the one
proposed by Wolfenstein~\cite{PhysRevLett.51.1945} in which the matrix
elements are expressed as powers of the parameter~$\lambda$
\begin{equation}
  \label{eq:II.13}
  \lambda\approx\lvert V_{12}\rvert^{2}.
\end{equation}
For 3~generations the CKM matrix in the Wolfenstein parameterization
has the form
\begin{equation}
  \label{eq:II.14}
 V^{(3)}=\left(
\begin{array}{ccc}
  1-\frac{\lambda^{2}}{2}&\lambda&A\lambda^{3}(\rho-i\eta)\\
-\lambda&1-\frac{\lambda^{2}}{2}&A\lambda^{2}\\
A\lambda^{3}(1-\rho-i\eta)&-A\lambda^{2}&1
\end{array}
\right).
\end{equation}
The parameterization given in Eq.~\eqref{eq:II.14} is approximate. It
can be given an exact meaning by assuming a one to one correspondence
between the Wolfenstein parameters $A$, $\lambda$, $\rho$  and $\eta$
and the parameters of the standard parameterization $s_{12}\equiv\lambda$, $s_{23}\equiv A\lambda^{2}$ and $s_{13}\text{e}^{i\delta_{13}}\equiv A\lambda^{3}(\rho+i\eta)$.~\cite{PhysRevD.50.3433}
We shall generalize the parameterization~\eqref{eq:II.14} to the case
of 4~quark generations using the method outlined earlier.

First we have to construct the vectors $\mathbf{e}_{i}$, which are expressed
in the spirit of the Wolfenstein parameterization in terms of the
powers of $\lambda$. The vectors $\mathbf{e}_{i}$ are real and are
chosen in the following way
\begin{equation}
\label{eq:II.15}
\begin{array}{llll}
\mathbf{e}_{1}=N_{1}\left( 
\begin{array}{c}
1+z_{2}^{2}+z_{3}^{2}\\
-z_{1}z_{2}\\
-z_{1}z_{3}\\
-z_{1}
\end{array}\right),

&\mathbf{e}_{2}=N_{2}\left( 
\begin{array}{c}
  0\\
  1+z_{3}^{2}\\
  -z_{2}z_{3}\\
  -z_{2}
\end{array}
\right),

\mathbf{e}_{3}=N_{3}\left( 
\begin{array}{c}
0\\
0\\
1\\
-z_{3}
\end{array}\right),

&\mathbf{e}_{4}=N_{4}\left( 
\begin{array}{c}
z_{1}\\
z_{2}\\
z_{3}\\
1
\end{array}\right) .

\end{array}
\end{equation}
Here $z_{i}=A_{i}^{(4)}\lambda^{k_{i}}$, $A_{i}^{(4)}\sim 1$,
$k_{i}\geq 1$ are integers and $N_{i}$ are suitable normalization
factors. The powers $k_{i}$ are considered to be constants (suppression factors) and $A_{i}^{(4)}$, $i=1,2,3$ are free parameters. It is easy to verify that the vectors $\mathbf{e}_{i}$ fulfill
Eq.~\eqref{eq:II.4a}, i.e., are orthogonal and normalized to~1.
The columns of the CKM matrix for 4~generations are then equal
\begin{equation}
  \label{eq:II.17}
  \begin{array}{l}
    \mathbf{v}_{1}^{(4)}= V_{11}^{(3)}\mathbf{e}_{1}
    +\text{e}^{-i\delta_{1,4}}V_{21}^{(3)}\mathbf{e}_{2}
    +\text{e}^{-i\delta_{2,4}}V_{31}^{(3)}\mathbf{e}_{3},\\
  \mathbf{v}_{2}^{(4)}= V_{12}^{(3)}\mathbf{e}_{1}
    +\text{e}^{-i\delta_{1,4}}V_{22}^{(3)}\mathbf{e}_{2}
    +\text{e}^{-i\delta_{2,4}}V_{32}^{(3)}\mathbf{e}_{3},\\
  \mathbf{v}_{3}^{(4)}= V_{13}^{(3)}\mathbf{e}_{1}
    +\text{e}^{-i\delta_{1,4}}V_{23}^{(3)}\mathbf{e}_{2}
    +\text{e}^{-i\delta_{2,4}}V_{33}^{(3)}\mathbf{e}_{3},\\
    \mathbf{v}_{4}^{(4)}=\mathbf{e}_{4},
  \end{array}
\end{equation}
and the CKM matrix is equal
\begin{equation}
  \label{eq:II.18}
  V^{(4)}=\left(\mathbf{v}_{1}^{(4)},\mathbf{v}_{2}^{(4)},
    \mathbf{v}_{3}^{(4)},\mathbf{v}_{4}^{(4)}\right).
\end{equation}
The matrix $V^{(4)}$ in Eq.~(\ref{eq:II.18}) is described by 9
parameters:
\begin{center}
  \begin{minipage}{0.8\linewidth}
$\lambda$, $A$, $\rho$, $\eta$ of the matrix~\eqref{eq:II.14}, introduced by Wolfenstein,\\
$A_{1}^{(4)}$, $A_{2}^{(4)}$, $A_{3}^{(4)}$, $\delta_{1,4}$,
$\delta_{2,4}$ of the vectors in~\eqref{eq:II.17}.
\end{minipage}
\end{center}
Not all these parameters can be determined from the experimental
data. On the other hand we can derive some restrictions on the powers
$k_{i}$ of the supression factors for the 4-th generation from the
experimental information for the CKM matrix for 3~generations. We have
the following information\footnote{If one uses unitarity of the
  $3\times 3$ CKM matrix then one has $(\lvert V_{12}\rvert -\lvert
  V_{21}\rvert)\sim\lambda^{5}$, but the element $\lvert V_{21}\rvert$
  is not measured with such a precision and experimentally we have
  $(\lvert V_{12}\rvert - \lvert V_{21}\rvert)\sim\lambda^{3}$ as
  in~\eqref{eq:II.19}.}
\begin{equation}
  \label{eq:II.19}
  \begin{array}{l}
    \lvert V_{12}\rvert\sim\lambda,\quad \lvert V_{21}\rvert\sim\lambda,\quad
    \lvert V_{23}\rvert\sim\lambda^{2},\quad \lvert
    V_{32}\rvert\sim\lambda^{2},\quad
\lvert V_{13}\rvert\sim\lambda^{3},\\  \lvert V_{31}\rvert\sim\lambda^{3},\quad
(\lvert V_{12}\rvert -\lvert V_{21}\rvert)\sim\lambda^{3},\quad
(\lvert V_{23}\rvert -\lvert V_{32}\rvert)\sim\lambda^{4}.
  \end{array}
\end{equation}
Now, using the information in Eq.~\eqref{eq:II.19} with the explicit
representation of the \mbox{$4\times 4$} CKM matrix in Eqs.~\eqref{eq:II.17}
and~\eqref{eq:II.18} we obtain the following restrictions on the
powers~$k_{i}$
\begin{equation}
  \label{eq:II.20}
  k_{i}\geq 1,\quad k_{1}+k_{2}\geq 3,\quad k_{2}+k_{3}\geq 4,\quad
  k_{1}+k_{3}\geq 3,
\end{equation}
which can be resolved and give
\begin{equation}
  \label{eq:II.21}
  k_{1}\geq 1,\quad k_{2}\geq 2,\quad k_{3}\geq 2.
\end{equation}
The vector $\mathbf{v}_{4}$ for the minimal values of $k_{i}$ in
Eq.~\eqref{eq:II.21} has the following form
\begin{equation}
  \label{eq:II.22}
  \mathbf{v}_{4}=N_{4}\left(
    \begin{array}{c}
      A_{1}^{(4)}\lambda\\
      A_{2}^{(4)}\lambda^{2}\\
      A_{3}^{(4)}\lambda^{2}\\
      1
    \end{array}
\right).
\end{equation}
This result is rather surprising, because in the case of 3 generations
the suppression has totally different structure. The full analysis of
the $4\times 4$ CKM matrix based on Eq.~\eqref{eq:II.18} will be
published elsewhere. However, we would like to note
that the real suppression may be different, because we have only
obtained the \emph{lower limits} of the suppression powers.

To conclude this section we will compare the values of two following rephasing invariants
\begin{equation}
  \label{eq:II.23A}
  \begin{array}{l}
    J_{A}=\operatorname{Im}(V_{12}V_{23}V_{13}^{*}V_{22}^{*})\\
    J_{B}=\operatorname{Im}(V_{21}V_{33}V_{23}^{*}V_{31}^{*})
  \end{array},
\end{equation}
that describe the CP violation effects in the strange and bottom
sectors. For 3~generations of quarks from the unitarity of the CKM
matrix it follows that $J_{A}=J$ and $J_{B}=-J$, where $J$ is the Jarlskog invariant. We thus have
\begin{equation}
  \label{eq:II.23}
  J_{A}+J_{B}=0, \text{ for 3 quark generations.}
\end{equation}
For 4~generations Eq.~\eqref{eq:II.23} does not hold and we have
\begin{equation}
  \label{eq:II.24}
  J_{A}+J_{B}\approx
  -\operatorname{Im}(\text{e}^{-i\delta_{2,4}}V_{21}^{(3)}
  V_{33}^{(3)}(V_{23}^{(3)}V_{11}^{(3)} )^{*}) z_{1}z_{3}.
\end{equation}
Experimentally $J_{A}$ is of the order $\lambda^{6}$ and $J_{B}$ is rather poorly known, because it contains the CKM matrix elements that are known with large errors. The sum $J_{A}+J_{B}$ gives the information how the CP violation effects differ in the strange and bottom sectors. If \mbox{$J_{A}+J_{B}=0$}, which holds exactly for 3~generations, then the CP violation parameters obtained from both sectors should be the same. If $J_{A}+J_{B}\sim\lambda^{6}$, then there is no cancellation between $J_{A}$ and $J_{B}$ and the CP violation in the strange and bottom sectors are not correlated. The estimated value of $J_{A}+J_{B}$ for 4~generations depends on the powers of the suppression factors through the sum $k_{1}+k_{3}$ and we obtain the following dependence of $J_{A}+J_{B}$ on this sum~\footnote{Note that according to Eqs.~\eqref{eq:II.20} and~\eqref{eq:II.21} we have $k_{1}+k_{3}\geq3$.}
\begin{equation}\label{eq:II.24CP}
\begin{minipage}{0.8\linewidth}
\begin{tabular}{c|c|l}
$k_{1}+k_{3}$&$J_{A}+J_{B}$\\
\hline
3&$\sim\lambda^{6}$&CP violation not correlated in the strange and\\
&&bottom sectors\\
4&$\sim\lambda^{7}$&CP violation weakly correlated in the strange\\
&&and bottom sectors (20\% difference)\\
5&$\sim\lambda^{8}$&CP violation strongly correlated in the strange\\
&&and bottom sectors (4\% difference)
\end{tabular}
\end{minipage}
\end{equation}

To conclude this section let us note that the presence of the 4-th
generation in the CKM matrix $V^{(3)}$ might be detected through the
violation of the unitarity of $V^{(3)}$. This can be done by
experimental verification of the asymmetries of the CKM matrix
\begin{equation}
  \label{eq:II.26}
  \lvert V_{12}\rvert^{2} -\lvert V_{21}\rvert^{2}=
  \lvert V_{23}\rvert^{2} -\lvert V_{32}\rvert^{2}=
  \lvert V_{31}\rvert^{2} -\lvert V_{13}\rvert^{2}
\end{equation}
which hold \emph{only} for 3~generations. Thus the experimental violation of Eq.~\eqref{eq:II.26} or observation that $J_{A}+J_{B}\neq 0 $ would be 
an experimental proof of the presence of 4-th generation.

\section{Rephasing monomials of the CKM matrix}\label{sec:3}

All observables of the CKM matrix are invariant under the rephasing of the quark fields. This invariance has important consequences on the properties of the CKM matrix and observables, like the reduction of the number of independent parameters of the CKM matrix. The rephasing invariant monomials built from the CKM matrix elements and its conjugates have been used in the discussion of various properties of the standard model related to the CKM matrix. The best known application of such a formalism is the Jarlskog's condition for CP symmetry~\cite{PhysRevD.36.2128}, and other applications also include the rephasing invariant parameterization of the CKM matrix and the discussion of the CP violation~\cite{Bjorken:1987tr,PhysRevLett.55.2935,PhysRevD.39.986,Jarlskog:2005zw,PhysRevLett.55.1039,PTP.92.289,Suzuki:2009my,PhysRevD.33.860}. 

In this section we will present the systematic study of the most general rephasing invariant monomials that can be built out of the CKM matrix elements and its conjugates. These monomials can be considered as building blocks of general observables of the CKM matrix. Next we will show that such invariant monomials can be expressed as powers of a finite number of elementary rephasing monomials.

The discussion of the rephasing invariant monomials depends on the number of generations and we will discuss here in detail the case of 3~generations. The generalization to 4~or more generations is in most cases simple, but may be tedious.

Let us denote by $P(m,n)$ the most general monomial constructed from the CKM matrix elements and its conjugates:
\begin{equation}
  \label{eq:III.1}
  P(m,n)=\Pi_{ij}(V_{ij})^{m_{ij}}\Pi_{kl}(V_{kl}^{*})^{n_{kl}}.
\end{equation}
Here $m$ and $n$ are $3\times 3$ matrices with integer~\footnote{The
  condition that the elements of the matrices $m$ and $n$ are integers
  may be relaxed, but the CKM observables are monomials that contain
  only integer powers.}, non negative matrix elements and
$[m]_{ij}=m_{ij}$ and $[n]_{ij}=n_{ij}$. The mapping between the
monomial $P(m,n)$ and the matrices $m$ and $n$ is one to
one~\footnote{For example, if the monomial $P(m,n)$ is equal to
\[
P(m,n)=V_{11}V_{22}V_{12}^{*}V_{21}^{*},
\]
then the matrices $m$ and $n$ are equal to
\[
m=
\left(
  \begin{array}{ccc}
    1&0&0\\
    0&1&0\\
    0&0&0
  \end{array}
\right),\quad
n=
\left(
  \begin{array}{ccc}
    0&1&0\\
    1&0&0\\
    0&0&0
  \end{array}
\right).
\]
}.

The monomials $P(m,n)$ fulfill the following properties
\begin{subequations}
  \label{eq:III.2}
  \begin{align}
    \label{eq:III.2a}
    &P(m_{1},n_{1})\cdot P(m_{2},n_{2})=P(m_{1}+m_{2},n_{1}+n_{2}),\\
    &(P(m,n))^{*}=P(n,m).
  \end{align}
\end{subequations}

The monomial $P(m,n)$ in general is not rephasing invariant. Suppose
that we make the following phase transformation of the CKM matrix
\begin{equation}
  \label{eq:III.3}
  V_{\text{CKM}}\rightarrow
  \operatorname{diag}(\text{e}^{i\phi_{1}},1,1) V_{\text{CKM}},
\end{equation}
then the monomial $P(m,n)$ is transformed in the following way
\begin{equation}
  \label{eq:III.4}
  P(m,n)\rightarrow \text{e}^{i\phi_{1}(m_{11}+m_{12}+m_{13}-n_{11}-n_{12}-n_{13})}
  P(m,n),
\end{equation}
so we see that $P(m,n)$ is invariant under the transformation in
Eq.~\eqref{eq:III.3} only if
\begin{equation}
  \label{eq:III.5}
  m_{11}+m_{12}+m_{13}=n_{11}+n_{12}+n_{13},
\end{equation}
i.e., if the sum of the elements of the first row of the matrices $m$
and $n$ are equal. From this one obtains
\begin{thm}\label{thm:1}
  The monomial $P(m,n)$ is rephasing invariant if the sums of the
  elements of the corresponding rows and columns of the matrices $m$
  and $n$ are equal. It means that for the rephasing invariant monomial $P(m,n)$ the matrices $m$ and $n$ fulfill the following conditions
  \begin{equation}
    \label{eq:III.6}
    \sum_{j=1}^{3}m_{ij}=\sum_{j=1}^{3}n_{ij},\quad 
    \sum_{j=1}^{3}m_{ji}=\sum_{j=1}^{3}n_{ji},\quad i=1,2,3.
  \end{equation}
\end{thm}

We want to consider here the rephasing invariants that carry the
information about the phases of the CKM matrix elements, e.g., the
Jarlskog rephasing invariant. The squares of the CKM matrix elements
$\lvert V_{ij}\rvert^{2}$ are rephasing invariant but they do not
contain any phase information and if we multiply any rephasing
invariant by $\lvert V_{ij}\rvert^{2k}$ then it does not change the phase
information in any way.  We therefore introduce the notion of the pure
rephasing invariant
\begin{definition}\label{def:1}
  The rephasing invariant monomial of the CKM matrix which cannot be factored
  out into the product of the absolute values of the elements of the
  CKM matrix and other invariant is called the \emph{pure rephasing
    invariant} (PRI).
\end{definition}
\begin{example} \textit{Rephasing invariant and pure rephasing invariant (PRI)}\\
The rephasing invariant
\[
V_{11}^{2}V_{22}V_{11}^{*}V_{12}^{*}V_{21}^{*}
\]
is not a \textit{pure rephasing invariant}, because it is equal to
\[
\lvert V_{11}\rvert^{2}V_{11}V_{22}V_{12}^{*}V_{21}^{*},
\]
i.e., one can factor out the square $\lvert V_{11}\rvert^{2}$ out of
it. On the other hand the remaining part
$V_{11}V_{22}V_{12}^{*}V_{21}^{*}$ is the pure rephasing invariant.
\end{example}

The PRIs can be represented by two matrices $m$ and
$n$, as in Eq.~(\ref{eq:III.1}), but they can also be represented by one $3\times
3$ matrix $p$ with the following properties:
\begin{enumerate}
\item The matrix elements of $p$ are integers (positive, negative or 0).
\item The sum of the elements of $p$ in each row and column is equal
  to 0.
  \item A permutation of the rows and columns of the $p$ matrix is reversible and the resulting matrix is also the $p$ matrix of pure rephasing invariant.
\end{enumerate}
The PRI, which we call $B(p)$, corresponding to the matrix $p$ is
constructed in the following way:
\begin{equation}
  \label{eq:III.7}
  B(p)=\Pi_{p_{ij}>0}(V_{ij})^{p_{ij}}\cdot\Pi_{p_{kl}<0}(V_{kl}^{*})^{-p_{kl}}.
\end{equation}
It is easy to show that $B(p)$ constructed in such a way is
rephasing invariant and  that it cannot be factored out into the
squares of the CKM matrix elements and other rephasing invariant, so
it is indeed the PRI.

The one to one mapping between the $p$ matrix and PRI $B(p)$ has the following additional properties
\begin{equation} \label{eq:III.7a}
\begin{split}
&(p_{1}+p_{2})\rightarrow B(p_{1}+p_{2})=B(p_{1})\cdot B(p_{2}),\quad n\cdot p\rightarrow (B(p))^{n},\quad n \text{ integer},\\
&\text{if } p\rightarrow B(p), \text{ then } (-p)\rightarrow (B(p))^{*}.
\end{split}
\end{equation}

\begin{example}\textit{Analytic expression of PRI for a given matrix $p$}\\
  Suppose that the matrix $p$ is equal
\[
p=\left(
  \begin{array}{rrr}
    1&-3&2\\
    2&1&-3\\
    -3&2&1
  \end{array}
\right)
\]
then the rephasing invariant defined by $p$ is equal
\[
V_{11}V_{13}^{2}V_{21}^{2}V_{22}V_{32}^{2}V_{33}(V_{12}^{3}V_{23}^{3}V_{31}^{3})^{*}
\]
and it fulfills all the properties of a PRI.
\end{example}

Let us introduce now the notion of the \emph{fundamental rephasing
  invariant} (FRI).
\begin{definition}
  The \emph{fundamental rephasing invariant (FRI)} is such a pure rephasing
  invariant monomial that is the product of 4~or 6~CKM matrix elements
  and its complex conjugates.
\end{definition}
It turns out that there are only 30~FRIs.  18~of them are products
of 4~CKM matrix elements and their complex conjugates and 12~are the
products of 6~CKM matrix elements and their complex conjugates. Their
explicit form is the following\\[3pt]
\emph{4-th order fundamental rephasing invariants ($J_{1},J_{2},\ldots,J_{18}$)}\footnote{Here, in the analogy to the Jarlskog invariant we use the same letter $J$ to denote the \mbox{4-th} order monomials, but note that the Jarlskog invariant contains the imaginary part and for monomials we do not take the imaginary part.}
\begin{equation}
  \label{eq:III.8}
  \begin{split}
  &J_{1}=V_{11}V_{22}V_{12}^{*}V_{21}^{*},\quad
  J_{5}=V_{11}V_{33}V_{13}^{*}V_{31}^{*},\\
  &J_{2}=V_{11}V_{23}V_{13}^{*}V_{21}^{*},\quad
  J_{6}=V_{12}V_{33}V_{13}^{*}V_{32}^{*},\\
  &J_{3}=V_{12}V_{23}V_{13}^{*}V_{22}^{*},\quad
  J_{7}=V_{21}V_{32}V_{22}^{*}V_{31}^{*},\\
  &J_{4}=V_{11}V_{32}V_{12}^{*}V_{31}^{*},\quad
  J_{8}=V_{21}V_{33}V_{23}^{*}V_{31}^{*},\\
  &J_{9}=V_{22}V_{33}V_{23}^{*}V_{32}^{*}\\
  &J_{9+i}=(J_{i})^{*},\quad i=1,\ldots,9.
 \end{split}
\end{equation}
\emph{6-th order fundamental rephasing invariants ($I_{1},I_{2},\ldots,I_{12}$)}
\begin{equation}
  \label{eq:III.9}
  \begin{split}
    &I_{1}=V_{11}V_{22}V_{33}V_{13}^{*}V_{21}^{*}V_{32}^{*},\quad
    I_{4}=V_{11}V_{23}V_{32}V_{13}^{*}V_{22}^{*}V_{31}^{*},\\
    &I_{2}=V_{11}V_{22}V_{33}V_{12}^{*}V_{23}^{*}V_{31}^{*},\quad
    I_{5}=V_{12}V_{23}V_{31}V_{13}^{*}V_{21}^{*}V_{32}^{*},\\
    &I_{3}=V_{11}V_{23}V_{32}V_{12}^{*}V_{21}^{*}V_{33}^{*},\quad
    I_{6}=V_{12}V_{21}V_{33}V_{13}^{*}V_{22}^{*}V_{31}^{*}\\
    &I_{6+i}=(I_{i})^{*},\quad i=1,\ldots,6.
  \end{split}
\end{equation}
For each FRI in Eqs.~\eqref{eq:III.8} and~\eqref{eq:III.9} there corresponds a $p$ matrix, e.g.,
\begin{equation}\label{eq:III.9a}
J_{1}\rightarrow p_{J_{1}}=\left(
\begin{array}{ccc}
1&-1&0\\
-1&1&0\\
0&0&0
\end{array}\right).\quad
I_{1}\rightarrow p_{I_{1}}=\left(
\begin{array}{ccc}
1&0&-1\\
-1&1&0\\
0&-1&1
\end{array}\right),\;\text{etc.}
\end{equation}
All the matrices $p_{i}$ corresponding to the invariants in Eqs.~\eqref{eq:III.8} and~\eqref{eq:III.9} can be obtained by the permutations of the rows and columns   the $p$ matrices of $J_{1}$ and $I_{1}$ that are given in Eq.~\eqref{eq:III.9a}. This means that an arbitrary permutation of the rows and columns of a $p_{J}$ matrix maps it into another $p_{J}$ matrix. The same applies to the $p_{I}$ matrices. The operation of the permutation of the rows and columns of the $p_{J}$ and $p_{I}$ matrices is reversible.

We can now formulate the following
\begin{thm}\label{thm:2}
  Any pure rephasing invariant can be expressed in a unique way as
  the product of positive powers of at most 4~fundamental rephasing
  invariants. Not more than one of these invariants can be from the
  set~\eqref{eq:III.9} and the remaining are from the set~\eqref{eq:III.8}.
\end{thm}
The important condition in Theorem~\ref{thm:2} is that the powers of the invariants are \emph{positive}. The next important point is that in the decomposition there may be no more than 1~fundamental rephasing invariant of the \mbox{6-th} order. Without these conditions the decomposition of a pure rephasing invariant into the fundamental invariants is not unique. The proof of this theorem is given in the Appendix.

The inverse theorem is not true, the product of two or more FRIs is
rephasing invariant, but it does not have to be the PRI.

From Theorem~\ref{thm:2} follows
\begin{thm}[Main Theorem for the Rephasing Invariants]\label{thm:main}
Any rephasing invariant monomial of the CKM matrix for 3~generations is the product of no more than 5~factors: 4~fundamental rephasing invariants taken to positive powers and the product of the squares of the absolute values of the CKM matrix elements also taken to positive powers. Only one fundamental invariant is from the set~\eqref{eq:III.9}.
\end{thm}

The Main Theorem has important consequences. From this theorem
follows that the imaginary part of any rephasing invariant monomial is
proportional to the Jarlskog invariant or equal to~0.

From the unitarity of the CKM matrix it follows that the \mbox{6-th} order FRIs in
Eq.~\eqref{eq:III.9} can be expressed by the \mbox{4-th} order FRIs from
Eq.~\eqref{eq:III.8} and the squares of the CKM matrix elements\footnote{It should be emphasized that without the unitarity of the CKM matrix there are no simple relations between the invariants of the 4-th and 6-th order. Thus relation~\eqref{eq:III.10} is also a test of the unitarity of the CKM matrix.}. We have for example,
\begin{equation}
  \label{eq:III.10}
  I_{1}=V_{11}V_{22}V_{33}V_{13}^{*}V_{21}^{*}V_{32}^{*} = \lvert
  V_{22}\rvert^{2}V_{12}V_{33}V_{13}^{*}V_{32}^{*} - \lvert
  V_{13}\rvert^{2}V_{22}V_{33} V_{23}^{*}V_{32}^{*}= \lvert
  V_{22}\rvert^{2} J_{6} - \lvert V_{13}\rvert^{2} J_{9}
\end{equation}
and there are analogous formulas for the remaining $I_{i}$'s.

To conclude this section let us briefly discuss some properties of the
rephasing invariants for 4~generations of quarks.

The notion of the fundamental rephasing invariant is generalized to
contain no more than 8~CKM matrix elements and there are 3~classes of
FRIs, with 4, 6 and 8~CKM matrix elements, respectively. The notion of
the pure rephasing invariant remains the same and the Main Theorem is
modified.

Let us also briefly discuss the unitarity properties of the FRIs with
4~CKM matrix elements. From simple considerations one can find out
that there are 36~such invariants (and its conjugates), e.g.,
\begin{equation}
  \label{eq:III.11}
  V_{11}V_{24}V_{14}^{*}V_{21}^{*}.
\end{equation}
Unitarity gives 48~relations between them, e.g.,
\begin{equation}
  \label{eq:III.12}
  \lvert V_{11}\rvert^{2} \lvert V_{21}\rvert^{2}
  +V_{12}V_{21}V_{11}^{*}V_{22}^{*} +V_{13}V_{21}V_{11}^{*}V_{23}^{*}
  +V_{14}V_{21}V_{11}^{*}V_{24}^{*} =0.
\end{equation}
If we take the imaginary parts of all these unitarity relations, then
we obtain 48~linear homogeneous equations for 36~variables. Not all
these equations are linearly independent and eventually it turns out
that only 9~of these equations are linearly independent. In
Ref.~\cite{PhysRevLett.55.1039} it has been shown that from unitarity of the CKM
matrix one can obtain further linear relations between these
9~imaginary parts and only 3~imaginary parts are sufficient to express
the remaining ones. The coefficients of the relations in the latter
step depend on the real parts of the invariants and the squares of the
CKM matrix elements, so the final expressions are complicated. We will
address this problem and discuss the consequences of the unitarity of
the CKM matrix for 4~generations elsewhere.

\section{Conclusions}\label{sec:4}
We have discussed here two important topics concerning the CKM matrix: parameterizations and the rephasing invariants of the CKM matrix.

From the theoretical point of view all exact parameterizations of the CKM matrix are equivalent. From the practical point of view the situation is less obvious, because certain experimental facts can be presented in a more transparent way in one parameterization than in the other. The scheme of the construction of the parameterization of the CKM matrix presented in this paper allows to adjust properties of the parameterization according to the needs. Such an approach has not been discussed before and it can facilitate the discussion of the properties of the CKM in the Standard Model or its extensions.

The next topic discussed in the paper was the rephasing invariance of the CKM matrix and the properties of the rephasing invariant monomials, built out of the CKM matrix. The Jarlskog invariant and unitarity triangle angles are examples of such monomials. Let us note that general rephasing invariant monomials constructed from the CKM matrix elements appear at higher orders of the renormalization group equations for the CKM matrix elements. The systematic analysis of such equations based on the results obtained in this paper will be published elsewhere.

The most important result concerning the rephasing invariance is stated in Theorem~\ref{thm:main} (Main Theorem) and it is mathematically a strong result. It tells that any rephasing invariant monomial of the CKM matrix can be expressed as the product of 5~factors which are functions of a finite, small number of the fundamental rephasing invariant monomials. The unitarity of the CKM matrix allows to express the \mbox{6-th} order rephasing invariant monomials by the \mbox{4-th} monomials and the \mbox{4-th} order monomials can be expressed by the squares of the absolute values of the CKM matrix element, but such an approach involves the subtractions of the terms that are very close (e.g., $(\lvert V_{12}\rvert^{2}-\lvert V_{12}\rvert^{2})\sim \lambda^{3}$), so it may lead to instability of the final result.

\begin{acknowledgments}
H.~P\'erez and P.~Kielanowski thank Professor Maria Krawczyk for hospitality in the Institute of Theoretical Physics at Warsaw University, Poland.
\end{acknowledgments}

\renewcommand{\theequation}{A.\arabic{equation}}
\setcounter{equation}{0}
\appendix\section*{Appendix: Proof of Theorem~\ref{thm:2}}
From Definition~\ref{def:1} and the discussion afterwords we know that there is one to one mapping between the \textit{pure rephasing invariants} and the $p$~matrices with the property that the sum
of the elements of each row and column is equal to~0. To prove the theorem we will analyze these matrices and we will show that the matrix corresponding to a PRI can be decomposed in a unique way as a linear combination with positive coefficients of at most 4 $p$~matrices corresponding to the fundamental rephasing invariants defined in Eqs.~\eqref{eq:III.8} and~\eqref{eq:III.9}. Then using the property~\eqref{eq:III.7a} of the $p$ matrices one obtains the Main Theorem.

The elements of a general $p$~matrix in~\eqref{eq:III.7} are positive or
negative integers and zeros. In~Table~\ref{tab:1} we list all possible
distributions of the number of the elements of the $p$ matrix which
are positive, negative or zero. The Type~3, 4, 6 and~7 contain two subtypes of distributions, which are related by complex conjugation and thus do not require separate proofs. As we know one can make permutations of rows and columns of the $p$ matrix and this operation has an inverse. In such a way we can simplify  the proof by organizing the elements of the $p$ matrix in the standard form without losing the generality.
\begin{table}[ht]
  \centering
  \begin{tabular}{c|c|c|cc|cc|c|cc|cc}
   Type&\phantom{A}1\phantom{A}&\phantom{A}2\phantom{A}&\multicolumn{2}{c|}{\phantom{A}3\phantom{A}}&\multicolumn{2}{c|}{\phantom{A}4\phantom{A}}&\phantom{A}5\phantom{A}&\multicolumn{2}{c|}{\phantom{A}6\phantom{A}}&\multicolumn{2}{c}{\phantom{A}7\phantom{A}}\\
\hline
   $k_{+}$&2&3&3&4&5&3&4&3&6&4&5\\
   $k_{-}$&2&3&4&3&3&5&4&6&3&5&4\\
   $k_{0}$&5&3&2&2&2&1&1&0&6&0&0
  \end{tabular}
  \caption{All possible distributions of the number of the elements of
    the $3\times 3$ $p$~matrix which are positive, negative or equal
    to~0. $k_{+}$ is the number of the elements that are positive,
    $k_{-}$ is the number of the elements that are negative and
    $k_{0}$ is the number of the elements that are equal to~0.}
  \label{tab:1}
\end{table}

We will consider now each type of the $p$ matrix listed in Table~\ref{tab:1}.\\
\textbf{Type~1.}
After a suitable permutation the $p$ matrix of the Type~1 has the following standard form
\begin{equation}\label{eq:A1} 
p^{1}=\left (
\begin{array}{ccc}
n&-n&0\\
-n&n&0\\
0&0&0
\end{array}
\right )=
n\times\left (
\begin{array}{ccc}
1&-1&0\\
-1&1&0\\
0&0&0
\end{array}\right )
,\quad n>0.
\end{equation}
Then from Eqs.~\eqref{eq:III.7a}, \eqref{eq:III.8} and~\eqref{eq:III.9a} one sees, that $p^{1}$ matrix corresponds to the $J_{1}$ fundamental invariant taken to the power $n$, that is
\begin{equation}\label{eq:A2}
p^{1}\rightarrow B(p^{1})=(J_{1})^{n}=(V_{11}V_{22}V_{12}^{*}V_{21}^{*})^{n}.
\end{equation}
\textbf{Type~2.} Here we have two types of nonequivalent $p$~matrices. The first one in the standard form reads
\begin{equation}\label{eq:A3}
p^{2_{A}}=\left (
\begin{array}{ccc}
n_{1}&n_{2}&-(n_{1}+n_{2})\\
-n_{1}&-n_{2}&(n_{1}+n_{2})\\
0&0&0
\end{array}\right )
=n_{1}\times\left (
\begin{array}{ccc}
1&0&-1\\
-1&0&1\\
0&0&0
\end{array}\right )+
n_{2}\times\left (
\begin{array}{ccc}
0&1&-1\\
0&-1&1\\
0&0&0
\end{array}\right ),\quad n_{1},n_{2}>0
\end{equation}
and $B(p^{2_{A}})$ has the representation
\begin{equation}\label{eq:A4}
p^{2_{A}}\rightarrow B(p^{2_{A}})=(J_{2})^{n_{1}}\cdot (J_{3})^{n_{2}}.
\end{equation}
The second nonequivalent $p$~matrix of the Type~2 has the form
\begin{equation}\label{eq:A3a}
p^{2_{B}}=\left (
\begin{array}{ccc}
n&0&-n\\
-n&n&0\\
0&-n&n
\end{array}\right )
=n\times\left (
\begin{array}{ccc}
1&0&-1\\
-1&1&0\\
0&-1&1
\end{array}\right ),\quad n>0
\end{equation}
and $B(p^{2_{B}})$ has the representation
\begin{equation}\label{eq:A5}
p^{2_{B}}\rightarrow B(p^{2_{B}})=(I_{1})^{n}.
\end{equation}
\textbf{Type~3.} The $p^{3}$ matrix in the standard form is equal
\begin{multline}\label{eq:A6}
p^{3}=\left (
\begin{array}{ccc}
n_{1}+n_{2}&0&-(n_{1}+n_{2})\\
-n_{2}&n_{2}&0\\
-n_{1}&-n_{2}&n_{1}+n_{2}
\end{array}\right )
=n_{1}\times\left (
\begin{array}{ccc}
1&0&-1\\
0&0&0\\
-1&0&1
\end{array}\right )
+n_{2}\times\left (
\begin{array}{ccc}
1&0&-1\\
-1&1&0\\
0&-1&1
\end{array}\right ),\\
n_{1},n_{2}>0
\end{multline}
and $B(p^{3})$ has the representation
\begin{equation}\label{eq:A7}
p^{3}\rightarrow B(p^{3})=(J_{5})^{n_{1}}\cdot (I_{1})^{n_{2}}.
\end{equation}
\textbf{Type~4.} The $p^{4}$ matrix after a suitable permutation of rows and columns takes the following standard form
\begin{multline}\label{eq:A8}
p^{4}=\left (
\begin{array}{ccc}
n_{1}&n_{2}&-(n_{1}+n_{2})\\
n_{3}&-(n_{1}+n_{2}+n_{3})&(n_{1}+n_{2})\\
-(n_{1}+n_{3})&(n_{1}+n_{3})&0
\end{array}\right )
=n_{1}\times\left (
\begin{array}{ccc}
1&0&-1\\
0&-1&1\\
-1&1&0
\end{array}\right )\\
+n_{2}\times\left (
\begin{array}{ccc}
0&1&-1\\
0&-1&1\\
0&0&0
\end{array}\right )
+n_{3}\times\left (
\begin{array}{ccc}
0&0&0\\
1&-1&0\\
-1&1&0
\end{array}\right ),\quad
n_{1},n_{2},n_{3}>0
\end{multline}
and $B(p^{4})$ has the representation
\begin{equation}\label{eq:A9}
p^{4}\rightarrow B(p^{4})=(I_{4})^{n_{1}}\cdot (J_{3})^{n_{2}}\cdot (J_{7})^{n_{3}}.
\end{equation}
\textbf{Type~5.} The standard form of the matrix $p^{5}$ is equal 
\begin{multline}\label{eq:A10}
p^{5}=\left (
\begin{array}{ccc}
n_{1}+n_{3}&n_{2}&-(n_{1}+n_{2}+n_{3})\\
-n_{3}&n_{3}&0\\
-n_{1}&-(n_{2}+n_{3})&n_{1}+n_{2}+n_{3}
\end{array}\right )
=n_{1}\times\left (
\begin{array}{ccc}
1&0&-1\\
0&0&0\\
-1&0&1
\end{array}\right )\\
+n_{2}\times\left (
\begin{array}{ccc}
0&1&-1\\
0&0&0\\
0&-1&1
\end{array}\right )
+n_{3}\times\left (
\begin{array}{ccc}
1&0&-1\\
-1&1&0\\
0&-1&1
\end{array}\right ),\quad
n_{1},n_{2},n_{3}>0
\end{multline}
and $B(p^{5})$ has the representation
\begin{equation}\label{eq:A11}
p^{5}\rightarrow B(p^{5})=(J_{5})^{n_{1}}\cdot (J_{6})^{n_{2}}\cdot (I_{1})^{n_{3}}.
\end{equation}
\textbf{Type~6.} The standard form of the matrix $p^{6}$ is equal 
\begin{multline}\label{eq:A12}
p^{6}=\left (
\begin{array}{ccc}
n_{1}+n_{3}+n_{4}&-(n_{1}+n_{4})&-n_{3}\\
-n_{1}&n_{1}+n_{2}+n_{4}&-(n_{2}+n_{4})\\
-(n_{3}+n_{4})&-n_{2}&n_{2}+n_{3}+n_{4}
\end{array}\right )
=n_{1}\times\left (
\begin{array}{ccc}
1&-1&0\\
-1&1&0\\
0&0&0
\end{array}\right )\\
+n_{2}\times\left (
\begin{array}{ccc}
0&0&0\\
0&1&-1\\
0&-1&1
\end{array}\right )
+n_{3}\times\left (
\begin{array}{ccc}
1&0&-1\\
0&0&0\\
-1&0&1
\end{array}\right )
+n_{4}\times\left (
\begin{array}{ccc}
1&-1&0\\
0&1&-1\\
-1&0&1
\end{array}\right ),\\
n_{1},n_{2},n_{3},n_{4}>0
\end{multline}
and $B(p^{6})$ has the representation
\begin{equation}\label{eq:A13}
p^{6}\rightarrow B(p^{6})=(J_{1})^{n_{1}}\cdot (J_{9})^{n_{2}}\cdot (J_{5})^{n_{3}}\cdot(I_{2})^{n_{4}}.
\end{equation}
\textbf{Type~7.} The standard form of the matrix $p^{7}$ reads
\begin{multline}\label{eq:A16}
p^{7}=\left (
\begin{array}{ccc}
n_{1}+n_{4}&n_{3}&-(n_{1}+n_{3}+n_{4})\\
-n_{4}&n_{2}+n_{4}&-n_{2}\\
-n_{1}&-(n_{2}+n_{3}+n_{4})&n_{1}+n_{2}+n_{3}+n_{4}
\end{array}\right )
=n_{1}\times\left (
\begin{array}{ccc}
1&0&-1\\
0&0&0\\
-1&0&1
\end{array}\right )\\
+n_{2}\times\left (
\begin{array}{ccc}
0&0&0\\
0&1&-1\\
0&-1&1
\end{array}\right )
+n_{3}\times\left (
\begin{array}{ccc}
0&1&-1\\
0&0&0\\
0&-1&1
\end{array}\right )
+n_{4}\times\left (
\begin{array}{ccc}
1&0&-1\\
-1&1&0\\
0&-1&1
\end{array}\right ),\\
n_{1},n_{2},n_{3},n_{4}>0
\end{multline}
and $B(p^{7})$ has the representation
\begin{equation}\label{eq:A17}
p^{7}\rightarrow B(p^{7})=(J_{5})^{n_{1}}\cdot (J_{9})^{n_{2}}\cdot (J_{6})^{n_{3}}\cdot(I_{1})^{n_{4}}.
\end{equation}
We have considered here all possible types of pure monomial rephasing invariants and thus Theorem~\ref{thm:2} follows from explicit calculation.  This completes the proof.


%

\end{document}